\documentclass[aps,preprint]{revtex4}
\usepackage{graphicx}
\usepackage{epsfig}
\usepackage{amsmath}
\usepackage{graphics}
\begin{document}

\title{Calculation of semiclassical free energy differences along non-equilibrium classical trajectories}

\author{M. F. Gelin}

\affiliation{Department of Chemistry, Technical University of Munich, Lichtenbergstrasse 4, D-85747 Garching, Germany}

\author{D. S. Kosov$^{1,2}$ }
\affiliation{
$^1$Department of Physics and  Center for Nonlinear Phenomena and Complex Systems,
Universit\'e Libre de Bruxelles, 
Campus Plaine, CP 231, 
Blvd du Triomphe, 
B-1050 Brussels, 
Belgium \\
$^2$Department of Chemistry and Biochemistry,
 University of Maryland, College Park, 20742, USA  }

\begin{abstract}
We have derived several relations, 
which allow the evaluation of the system free energy changes in the leading order in $\hbar^{2}$
along classically generated trajectories.
The  results are formulated in terms of  purely
classical Hamiltonians and trajectories, so that semiclassical partition
functions can be computed, e.g., via classical molecular dynamics
simulations. The Hamiltonians, however, contain additional potential-energy
terms, which are proportional to $\hbar^{2}$ and are temperature-dependent.
We discussed the influence of quantum interference on the nonequilibrium work and 
problems with unambiguous definition of the semiclassical work operator.
\end{abstract}

\maketitle

\section{Introduction}

The nonequilibrium work theorem, or Jarzynski equation, relates nonequilibrium
work performed on a dissipative system during certain time interval
to the difference of the corresponding equilibrium free energies. Formulated
first for classical systems \cite{kuz81a,jar97,jar04}, the work theorem
has been generalized to the quantum case, too \cite{kuz81a,gaspard,yuk00,Htas00,kurchan01,
muk03,muk06,nie05,han07,lutz08,nol07,mahler07,mae04,mon05,kosJa, han09, gasp08,cro08}.
Monnai and Tasaki have found quantum ($\sim \hbar^{2}$) corrections to the transient 
and steady-state fluctuation theorems for a damped harmonic oscillator \cite{MT}. 
Chernyak and Mukamel gave  quantum ($\sim \hbar^{2}$) corrections to the work theorem 
for a specific form of the driven system Hamiltonian \cite{Chernyak04}.  
The quantum generalizations of the work theorem have triggered a certain
controversy in the literature on whether it is possible to uniquely
define the quantum work operator and how to interpret the quantum
work theorem \cite{nie05,han07,lutz08,nol07,mahler07,kosJa,han09}. 

In the present paper, we use the ideas inspired by  the classical work theorems to derive 
several relations which allow us the calculation of equilibrium semiclassical 
free energy changes along non-equilibrium classical trajectories. 
Our aim is twofold. First, we develop a practical tool for the calculation of
semiclassical free energies differences through  classical
equations of motion and/or molecular dynamics simulations. 
Such semiclassical corrections are responsible, e.g., for the short-distance anomaly 
of the pair distribution function of liquid neon \cite{powles83,zoppi85}.   
Second, we  wish to get a better understanding
of how quantum interference influences the nonequilibrium work 
and  if it is possible to uniquely define the semiclassical work
operator.


Note that all the semiclassical quantities which are considered in
the present paper are evaluated in the leading order in $\hbar^{2}$,
i.e., with the accuracy $O(\hbar^{4})$. 
We use  the simbols $h$ and $\hbar$ intermittently throughout the text, 
in order to avoid writing numerous factors of $2\pi$.

\section{Wigner distributions and useful identities}

Let us consider a collection of $N$ quantum point particles with
the Hamiltonian 

\begin{equation}
\hat{H}(\hat{\mathbf{p}},\hat{\mathbf{q}})=
\sum_{i=1}^{N}\frac{\hat{p}_{i}^{2}}{2m_{i}}+U(\hat{\mathbf{q}}).\label{H}\end{equation}
$\hat{q}_{i}$, $\hat{p}_{i}=-i\hbar d/d\hat{q}_{i}$, and $m_{i}$
are the positions, momenta, and masses of the particles, 
$U(\hat{\mathbf{q}})$
is the potential energy. Hereafter, the boldface notation 
$\hat{\mathbf{p}},$ $\hat{\mathbf{q}}$
is used to collectively denote the set of all $\hat{p}_{i}$ and $\hat{q}_{i}$.
We assume that the particles are prepared in an equilibrium canonical ensemble
at the temperature $T.$ The corresponding distribution (density matrix)
reads \begin{equation}
\hat{\rho}(\hat{\mathbf{p}},\hat{\mathbf{q}})=Z^{-1}\exp\{-\beta\hat{H}(\hat{\mathbf{p}},\hat{\mathbf{q}})\},\,\,\, Z=\textrm{Tr}(\exp\{-\beta\hat{H}(\hat{\mathbf{p}},\hat{\mathbf{q}})\}),\label{ro}\end{equation}
$Z$ being the partition function.

We wish to calculate the leading ($\sim\hbar^{2}$) corrections to
the canonical distribution (\ref{ro}) and partition function $Z$.
To this end, it is convenient to switch to the Wigner representation
\cite{wig32,wig84}, so that (\ref{ro}) becomes the corresponding
Wigner distribution (the subscript $W$) 
\begin{equation}
\rho_{W}^{(0)}(\mathbf{p},\mathbf{q})=Z^{-1}\exp\{-\beta(H(\mathbf{p},\mathbf{q})+h^{2}\Delta^{(0)}(\beta,\mathbf{p},\mathbf{q}))\}+O(h^{4}),\label{rWgen}\end{equation}
\begin{equation}
Z=\int d\mathbf{p}d\mathbf{q}\exp\{-\beta(H(\mathbf{p},\mathbf{q})+h^{2}\Delta^{(0)}(\beta,\mathbf{p},\mathbf{q}))\}.\label{ZWgen}\end{equation}
Here $\mathbf{p}$ and $\mathbf{q}$ can be treated as the phase variables
in the Wigner space, taking the trace reduces to the integration over
the Wigner phase space variables, $H(\mathbf{p},\mathbf{q})$ is the
classical Hamiltonian corresponding to its quantum counterpart (\ref{H}),
 and $\Delta^{(0)}$
is the temperature-dependent quantum correction \cite{wig84,wig32}
\begin{equation}
\Delta^{(0)}=\sum_{i=1}^{N}\left\{ \frac{\beta}{8m_{i}}\frac{\partial^{2}U(\mathbf{q})}{\partial q_{i}^{2}}-\frac{\beta^{2}}{24m_{i}}\left(\frac{\partial U(\mathbf{q})}{\partial q_{i}}\right)^{2}\right\} -\sum_{i,j=1}^{N}\frac{\beta^{2}p_{i}p_{j}}{24m_{i}m_{j}}\frac{\partial^{2}U(\mathbf{q})}{\partial q_{i}\partial q_{j}}.\label{W0}\end{equation}

For our further purposes we preaverage $\Delta^{(0)}$ over momenta.
That is, we replace $p_{i}p_{j}$ by $\delta_{ij}m_{i}/\beta$ and
obtain \begin{equation}
\Delta^{(1)}=\sum_{i=1}^{N}\left\{ \frac{\beta}{12m_{i}}\frac{\partial^{2}U(\mathbf{q})}{\partial q_{i}^{2}}-\frac{\beta^{2}}{24m_{i}}\left(\frac{\partial U(\mathbf{q})}{\partial q_{i}}\right)^{2}\right\} .\label{W1}\end{equation}
We also introduce two new functions \begin{equation}
\Delta^{(2)}=\sum_{i=1}^{N}\left\{ \frac{\beta}{24m_{i}}\frac{\partial^{2}U(\mathbf{q})}{\partial q_{i}^{2}}\right\} ,\label{W2}\end{equation}
\begin{equation}
\Delta^{(3)}=\sum_{i=1}^{N}\left\{ \frac{\beta^{2}}{24m_{i}}\left(\frac{\partial U(\mathbf{q})}{\partial q_{i}}\right)^{2}\right\}. \label{W3}\end{equation}
The physical meaning of these new functions will be explained below.
Let us now consider the quantities \begin{equation}
\rho_{W}^{(a)}(\mathbf{p},\mathbf{q})=Z^{-1}\exp\{-\beta(H(\mathbf{p},\mathbf{q})+h^{2}\Delta^{(a)}(\beta,\mathbf{p},\mathbf{q}))\},\,\,\, a=0,1,2,3.\label{rWi}
\end{equation}
Here
$\rho_{W}^{(0)}$ is the semiclassical Wigner distribution (\ref{rWgen}) but 
$\rho_{W}^{(a)}$ for $a=1,2,3$ are not true semiclassical Wigner
distributions \cite{foot6}. However, all  $\rho_{W}^{(a)}$ 
share the following evident property: The partition
functions\begin{equation}
Z^{(a)} = \int d\mathbf{p}d\mathbf{q}\exp\{-\beta(H(\mathbf{p},\mathbf{q})
+h^{2}\Delta^{(a)}(\beta,\mathbf{p},\mathbf{q}))\} 
\label{Zi}\end{equation}
coincide with the true semiclassical partition function $Z^{(0)}$ 
within the accuracy $O(h^{4})$. The equivalence of $Z^{(1)}$
and $Z^{(2)}$, $Z^{(3)}$ can be demonstrated via integration by parts,\[
\beta\int d\mathbf{q}\exp\{-\beta U(\mathbf{q})\}\left(\frac{\partial U(\mathbf{q})}{\partial q_{i}}\right)^{2}=\int d\mathbf{q}\exp\{-\beta U(\mathbf{q})\}\frac{\partial^{2}U(\mathbf{q})}{\partial q_{i}^{2}}.\]


\section{Semiclassical free energy changes along classical trajectories}


We are in a position now to derive the semiclassical version of the
nonequilibrium work theorem. Let 
\begin{equation}
\hat{H}(\hat{\mathbf{p}},\hat{\mathbf{q}},t)=\sum_{i=1}^{N}\frac{\hat{p}_{i}^{2}}{2m_{i}}+U(\hat{\mathbf{q}},t).\label{Ht}\end{equation}
 be a quantum Hamiltonian, which is allowed to be explicitly time-dependent. Let us now introduce the quantities \begin{equation}
H^{(a)}(\beta,\mathbf{p},\mathbf{q},t)= H(\mathbf{p},\mathbf{q},t)+h^{2}\Delta^{(a)}(\beta,\mathbf{p},\mathbf{q},t),\,\, a=0,1,2,3.\label{Hi}\end{equation}
Here $H(\mathbf{p},\mathbf{q},t)$ is the classical Hamiltonian which
corresponds to the quantum Hamiltonian (\ref{Ht}), and functions
$\Delta^{(a)}(\beta,\mathbf{p},\mathbf{q},t)$ are constructed out
of $H(\mathbf{p},\mathbf{q},t)$ as is prescribed via Eqs. (\ref{W0})-(\ref{W3}). 

Let us now consider $H^{(a)}$ for $a=1,2,3$ as 
{\it classical} Hamiltonians
which are parametrically temperature-dependent \cite{foot4}.   They can equivalently
be rewritten as \begin{equation}
H^{(a)}(\beta,\mathbf{p},\mathbf{q},t) = \sum_{i=1}^{N}\frac{p_{i}^{2}}{2m_{i}}+U^{(a)}(\beta,\mathbf{q},t),\,\,\, U^{(a)}(\beta,\mathbf{q},t)= U(\mathbf{q},t)+h^{2}\Delta^{(a)}(\beta,\mathbf{q},t).\label{Hia}
\end{equation}
So, the quantum correction gives rise to a simple redefinition of
the potential energy, which acquires an additional temperature-dependent
$\sim\hbar^{2}$ contribution. 
Now we can use the 
Hamiltonian (\ref{Hia}) to  write down the corresponding
classical equations of motion \begin{equation} 
\frac{dp_{i}(t)}{dt}=-\frac{H^{(a)}(\beta,\mathbf{p},\mathbf{q},t)}{dq_{i}},\,\,\,\frac{dq_{i}(t)}{dt}=
\frac{H^{(a)}(\beta,\mathbf{p},\mathbf{q},t)}{dp_{i}}\label{Em}\end{equation}
which produce the classical trajectories $\mathbf{p}(t),\,\mathbf{q}(t)$.
Furthermore, we can treat \begin{equation}
\rho_{W}^{(a)}(\mathbf{p},\mathbf{q},t)=Z_{t}^{-1}\exp\{-\beta H^{(a)}(\beta,\mathbf{p},\mathbf{q},t)\}\label{Wt}\end{equation}
 as the corresponding classical canonical distribution with the partition
function\begin{equation}
Z_{t}=\int d\mathbf{p}d\mathbf{q}\exp\{-\beta H^{(a)}(\beta,\mathbf{p},\mathbf{q},t)\}.\label{Zt}\end{equation}
 After this is done, we can proceed analogously to the derivation
of the classical Jarzynski equation \cite{jar04,kosJa}. Namely, we
can write the following chain of identities:\[
\int d\mathbf{p}_{0}d\mathbf{q}_{0}\exp\{-\beta H^{(a)}(\beta,\mathbf{p}_{0},\mathbf{q}_{0},0)\}\exp\{-\beta(H^{(a)}(\beta,\mathbf{p}_{t},\mathbf{q}_{t},t)-H^{(a)}(\beta,\mathbf{p}_{0},\mathbf{q}_{0},0))\}=\]
 \begin{equation}
\int d\mathbf{p}_{0}d\mathbf{q}_{0}\exp\{-\beta H^{(a)}(\beta,\mathbf{p}_{t},\mathbf{q}_{t},t)\}=\int d\mathbf{p}_{t}d\mathbf{q}_{t}\exp\{-\beta H^{(a)}(\beta,\mathbf{p}_{t},\mathbf{q}_{t},t)\}= Z_{t},\label{AB1}\end{equation}
$a=1,2,3$. In deriving Eq. (\ref{AB1}) we make use of the fact that
motion of a Hamiltonian system can be regarded as the canonical transformation,
for which the Liouville theorem holds: $d\mathbf{p}_{0}d\mathbf{q}_{0}$  $=d\mathbf{p}_{t}d\mathbf{q}_{t}$.
Introducing the abbreviation \begin{equation}
\int d\mathbf{p}_{0}d\mathbf{q}_{0}Z_{0}^{-1}\exp\{-\beta H^{(a)}(\beta,\mathbf{p}_{0},\mathbf{q}_{0},0)\}...=\left\langle ...\right\rangle_{0} \label{av}\end{equation}
and dividing Eq. (\ref{AB1}) by $Z_{0}$, we obtain
\begin{equation}
\left\langle \exp\{-\beta(H^{(a)}(\beta,\mathbf{p}_{t},\mathbf{q}_{t},t)-
H^{(a)}(\beta,\mathbf{p}_{0},\mathbf{q}_{0},0))\}\right\rangle_{0} =Z_{t}/Z_{0},\,\,\, a=1,2,3.\label{AB2}\end{equation}

We can further split the total Hamiltonian into the system Hamiltonian,
the bath Hamiltonian and their coupling, \begin{equation}
H^{(a)}(\beta,\mathbf{p},\mathbf{q},t)=H_{S}^{(a)}(\beta,\mathbf{p}_{S},\mathbf{q}_{S},t)
+H_{B}^{(a)}(\beta,\mathbf{p}_{B},\mathbf{q}_{B})+
H_{SB}^{(a)}(\beta,\mathbf{p}_{S},\mathbf{q}_{S},\mathbf{p}_{B},\mathbf{q}_{B}).\label{SB}\end{equation}
Here the system Hamiltonian only is allowed to be explicitly time-dependent.
Plugging $H^{(a)}$ (\ref{SB}) into the identity (\ref{AB2}) and
making use of the equations of motion (\ref{Em}), we can write\[
H^{(a)}(\beta,\mathbf{p}_{t},\mathbf{q}_{t},t)-H^{(a)}(\beta,\mathbf{p}_{0},\mathbf{q}_{0},0)=\]
\begin{equation}
\int_{0}^{t}dt'\frac{d}{d t'}H^{(i)}(\beta,\mathbf{p}(t'),\mathbf{q}(t'),t')=\int_{0}^{t}dt'\frac{\partial}{\partial t'}H_{S}^{(i)}(\beta,\mathbf{p}_{S}(t'),\mathbf{q}_{S}(t'),t')= W^{(a)},\label{Ja1}\end{equation}
$W^{(a)}$ being the work performed on our {\it classical} system. Thus Eqs. (\ref{AB2})
and (\ref{Ja1}) yield 
\begin{equation}
\left\langle \exp(-\beta W^{(a)})\right\rangle_{0} =Z_{t}/Z_{0},\label{Ja0}\end{equation}
which is the semiclassical analogue of the classical Jarzynski formula \cite{jar04}.

Note that the ratio of the total ($S+B$) partition functions $Z_{t}/Z_{0}$ equals to the ratio of the system partition functions \cite{jar04}. Indeed,
the reduced system ($S$) density matrix $\rho_{S}^{(a)}$ is defined
via averaging the total density matrix (\ref{Wt}) over the bath
degrees of freedom \cite{roux,MM}:
\begin{equation}
\rho_{S}^{(a)}(\mathbf{p}_{S},\mathbf{q}_{S},t)=Z_{S,t}^{-1}\int d\mathbf{p}_{B}d\mathbf{q}_{B}\exp\{-\beta(H^{(a)}(\beta,\mathbf{p}_{S},\mathbf{q}_{S},\mathbf{p}_{B},\mathbf{q}_{B})-F_{B})\},\label{WtS}\end{equation}
 where the bath free energy $F_{B}=-(\ln Z_{B})/\beta$ is determined
through the bath partition function\begin{equation}
Z_{B}=\int d\mathbf{p}_{B}d\mathbf{q}_{B}\exp\{-\beta H_{B}^{(a)}(\beta,\mathbf{p}_{B},\mathbf{q}_{B})\}.\label{ZB}\end{equation}
The system partition function in Eq. (\ref{WtS}) is explicitly
defined as $Z_{S,t}=Z_{t}/Z_{B}$ and we can write
\begin{equation}
Z_{t}/Z_{0}=Z_{S,t}/Z_{S,0}.\label{ZZS}\end{equation}
If the quantum $\sim\hbar^{2}$ corrections are neglected, then Eqs.
(\ref{AB2}) and (\ref{Ja0}) reduce, of course, to the classical work
theorem for Hamiltonian systems \cite{jar04}. 

To illustrate computational aspects of practical use of relations (\ref{AB2},\ref{Ja0}), we performed "molecular dynamics" simulations for a point particle in 
one-dimensional nonlinear potential. The quantum Hamiltonian reads  
\begin{equation}
H(\hat{p},\hat{q},t)=\frac{\hat{p}^{2}}{2m}+a\hat{q}^{2} + b\hat{q}^{4}\frac{t}{1+t}.\label{Ham}\end{equation}
In this example, the mass $m=1$, the parameters $a=b=1$, the position  $\hat{q}$, the momentum 
$\hat{p}=-i\hbar d/d\hat{q}$,  the time $t$, and the Planck constant are taken as dimensionless. 
We construct the corresponding semiclassical Hamiltonian $H^{(3)}(\beta,p,q,t)$ 
as is explained in Sections II and III. Then we calculate the free energy difference at times $t$ and $0$, 
$\Delta F = F_{t}-F_{0} = -\ln (Z_{t}/Z_{0})/ \beta$, by "molecular dynamics" simulations and "exactly". In "molecular dynamics" simulations, we sample initial values of $q$ and $p$ according to the  Wigner distribution (\ref{Wt}) ($N_{sam}$ being the number of samplings), solve equations of motions (\ref{Em}) numerically, and calculate the quantity $\exp \{-\beta (H^{(3)}(\beta,p,q,t)-H^{(3)}(\beta,p,q,0))\}$. The procedure is repeated $N_{sam}$ times, the averaged value  
$\left\langle \exp\{-\beta (H^{(3)}(\beta,p,q,t)-H^{(3)}(\beta,p,q,0))\}\right\rangle_{0}$ 
is obtained, and  $\Delta F$ is finally calculated through the semiclassical work theorem (\ref{AB2}). "Exact" $\Delta F$ is obtained trough  the direct numerical evaluation of the semiclassial partition functions $Z_{t}$ and $Z_{0}$  according to Eq. (\ref{Zt}). The results of the calculations are depicted in figure 1. We see that making $N_{sam}=10^2$ samplings gives already reasonable, but quite noisy estimation for $\Delta F$ (dashed lines), $N_{sam}=10^3$ gets a better result (dotted lines).  
$N_{sam}=10^4$ yields $\Delta F$  which is virtually indistinguishable with the 
exact results (full lines). The convergence is thus rather slow (see Ref. \cite{zucker06}  for the comparison of different simulation schemes for obtaining free energy differences). On the other hand, the convergence speed is the same for  classical (bottom curves, $h=0$) and semiclassical (upper curves, $h=1$) case \cite{foot1}.

The Hamiltonian $H^{(0)}(\beta,\mathbf{p},\mathbf{q})$, $a=0$,  has already been used in 
molecular dynamics simulations \cite{ss}. Such a choice gives a correct semiclassical distribution, but its practical implementation may encounter certain difficulties \cite{foot4,ss}. 
The choice $a=1$ gives a correct semiclassical distribution over positions $\mathbf{q}$.
The corresponding Hamiltonian $H^{(1)}(\beta,\mathbf{p},\mathbf{q})$ has also found its application in molecular dynamics simulations, being evaluated 
up to the terms of the order of $\hbar^2$ \cite{powles83} and  $\hbar^6$  \cite{zoppi85,zoppi87}.
To our knowledge, the Hamiltonians $H^{(2)}(\beta,\mathbf{p},\mathbf{q})$ 
and $H^{(3)}(\beta,\mathbf{p},\mathbf{q})$ have never been used in practice. However, they give 
more tractable expressions for the semiclassical potential $\Delta^{(a)}(\mathbf{q},t)$. 
It should be noted that (almost) all physically relevant potentials $U(\mathbf{q})\rightarrow 0$ when $|\mathbf{q}|\rightarrow \infty$. In that case, there is no problem with the exponentiating the semiclassical correction potentials, e.g., 
\begin{equation}
\exp \{-\beta H(\mathbf{p},\mathbf{q},t)\}
(1+h^{2}\beta\Delta^{(a)}(\beta,\mathbf{p},\mathbf{q},t)) =  
\exp \{-\beta H^{(a)}(\beta,\mathbf{p},\mathbf{q},t)\} +O(h^{4}).\label{Hex}\end{equation}
Such exponentiation may work even beyond its strict domain of validity $O(\hbar^4)$, since it corresponds to the partial summation of the higher order contributions in the semiclassical expansion. 
If we use  strongly attractive potentials, then exponentiation in Eq. (\ref{Hex}) for $a=1$ may not be feasible, since the contribution due to the second term in Eq. (\ref{W1}) may become predominant, 
so that the exponential explodes. In such a case, 
the choices $a=2$ and $a=3$ are necessary. 
Finally, a word of caution concerning applications of the present (and similar) 
semiclassical methods to  realistic molecular systems. 
The ubiquitous presence of sharply varying repulsions
in condensed phases means that the terms like (\ref{W2}) and (\ref{W3}) 
can be quite large. This can lead to numerical difficulties  \cite{ss}. 
More fundamentally, this  could indicate  that a straightforward  $\hbar^{2}$-perturbative
treatment might not always be appropriate.

\section{Difficulties with the definition of semiclassical work}

Several important points are to be discussed here in connection with the
interpretation of our results as semiclassical work theorems. 
As has been stressed in Sec. II, the distributions $\rho_{W}^{(a)}(\mathbf{p},\mathbf{q})$
(\ref{rWi}) for $a=1,2,3$ are not true semiclassical Wigner distributions.
However, they give correct values of the semiclassical partition functions
within the accuracy $O(h^{4})$. Furthermore, the Hamiltonian
dynamics governed by Eqs. (\ref{Em}) is not, of course,
the true semiclassical dynamics. However, if we run classical
molecular dynamics simulations with the Hamiltonians $H^{(a)}(\beta,\mathbf{p},\mathbf{q},t)$ ($a=1, 2, 3$) 
we get correct semiclassical values of the partition functions, again
with the accuracy $O(h^{4})$. 

The explicit form of the semiclassical nonequilibrium "work theorem"
(Eqs. (\ref{AB2}) and (\ref{Ja0})) is not unique. We have simultaneously
derived four ($a=0,1,2,3$) different "work theorems" with different
definitions of the work operator $W^{(a)}$ (\ref{Ja1}). For each particular 
choice of the classical Hamiltonian,  
$W^{(a)}$ can definitely be treated as the  classical work, and 
 all $W^{(a)}$ coincide in the limit $\hbar \rightarrow 0$.
However, we cannot regard any of $W^{(a)}$ as a true semiclassical work operator. 
First, $W^{(a)}$ for $a=0,1,2,3$ yield  different values of work, 
and it is not clear which choice (if any) is preferable. 
Second, the mean value of the work operator $\overline  W^{(a)}$  does not coincide 
with the mean semiclassical energy difference $\Delta E$. 
Indeed, adopting the notation  
\begin{equation}
\int d\mathbf{p}_{t}d\mathbf{q}_{t}Z_{t}^{-1}\exp\{-\beta H^{(a)}(\beta,\mathbf{p}_{t},\mathbf{q}_{t},t)\}...=\left\langle ...\right\rangle_{t}, \label{av1}\end{equation}
we can explicitly define  
\begin{equation}
\overline  W^{(a)} 
= 
\left\langle H^{(a)}(\beta,\mathbf{p}_{t},\mathbf{q}_{t},t)\right\rangle_{t}-
\left\langle H^{(a)}(\beta,\mathbf{p}_{0},\mathbf{q}_{0},0)\right\rangle_{0}
\label{Wav}\end{equation}
and 
\begin{equation}
\Delta E = 
\left\langle H(\beta,\mathbf{p}_{t},\mathbf{q}_{t},t)\right\rangle_{t}-
\left\langle H(\beta,\mathbf{p}_{0},\mathbf{q}_{0},0)\right\rangle_{0}.
\label{Eav}\end{equation}
Therefore,
\begin{equation}
\overline  W^{(a)}= \Delta E  + h^{2}\left\{
\left\langle \Delta^{(a)}(\beta,\mathbf{q}_{t},t)\right\rangle_{t}-
\left\langle \Delta^{(a)}(\beta,\mathbf{q}_{0},0)\right\rangle_{0}\right\}.\label{DE}\end{equation}
The difference between $W^{(a)}$ and $\Delta E$ is of the order of $h^{2}$ and is generally nonzero \cite{foot8}. This is the semiclassical way of stating that it is hardly  possible to properly 
and unambiguously define the quantum work operator \cite{nie05,han07,lutz08,nol07,mahler07,kosJa,han09}. 

It is important that $W^{(a)}$ (\ref{Ja1}) is explicitly
temperature dependent. That is a direct consequence of quantum interference.
In classical case, we can consider a single trajectory in the phase
space and calculate the work along it. Such a work is determined exclusively
by the difference of energies along the trajectory. Incoherent summation
over many such trajectories, weighted via classical canonical distribution,
yields the classical work theorem. In quantum or even semiclassical
case, there exists interference between the trajectories, which manifests
itself through the temperature dependent Hamiltonians $H^{(a)}(\beta,\mathbf{p},\mathbf{q},t)$. Thus, even in the semiclassical limit,
it is not possible to define $W^{(a)}$ as a function of exclusively
the phase space variables and time. The ensemble quantity,
viz., the temperature, inevitably enters the definition of work. 


The semiclassical Jarzynski equations (\ref{AB2}) and (\ref{Ja0})
cannot be directly considered as semiclassical versions of the quantum
Jarzynski equation \cite{kuz81a,gaspard,yuk00,Htas00,kurchan01,
muk03,muk06,nie05,han07,lutz08,nol07,mahler07,mae04,mon05,kosJa,han09, gasp08,cro08}. Indeed, the latter is written as \cite{han07}\begin{equation}
\left\langle \exp(\beta\hat{H}(0))\exp(-\beta\hat{H}(t))\right\rangle =\mathcal{T}_{<}\left\langle \exp(-\beta\hat{W})\right\rangle =Z_{t}/Z_{0}.\label{YaQQ}\end{equation}
Here $\hat{H}(t)$ is the total  Hamiltonian in the Heisenberg representation,
$\left\langle ...\right\rangle =\textrm{Tr}(\exp\{-\beta\hat{H}(0)\}...)$,
and $\mathcal{T}_{>}$ is the chronological ordering operator. The
trajectory-dependent work operator is explicitly defined as \cite{kosJa,han09}
\begin{equation}
\hat{W}=\hat{H}(t)-\hat{H}(0)=\int_{0}^{t}dt'\frac{d}{d t'}\hat{H}(t')=\int_{0}^{t}dt'\frac{\partial}{\partial t'}\hat{H}_{S}(t')\label{WQ}\end{equation}
(the last expression in Eq. (\ref{WQ}) holds true provided we partition
the total Hamiltonian as a sum of the system Hamiltonian, bath Hamiltonian
and their coupling, and assume that the system Hamiltonian is explicitly
time dependent, $H(t)=H_{S}(t)+H_{B}+H_{SB}$).
In principle, we can write down a direct semiclassical analogue of Eq. (\ref{YaQQ}).
The corresponding formulas are not presented here, since they are 
quite cumbersome and difficult to interpret. What is even more
important, we have to introduce the semiclassical time-evolution operator,
which should be evaluated in the leading order in $h^{2}$. That would
prevent us from a simple classical interpretation of the formulas.


\section{Conclusions}

We have derived semiclassical analogues (\ref{AB2}) and (\ref{Ja0}) of the classical work theorems,
which allow us to evaluate the ratio of the (system) partition functions
in the leading order in $\hbar^{2}$, with the accuracy $O(\hbar^{4})$. 
The corrections due to the quantum statistics (Bose-Einstein or Fermi-Dirac)
give rise to the contributions $\sim\hbar^{3}$ to the equilibrium
density matrices and partition functions \cite{LL5} and are not considered
here.

The semiclassical analogues of the work  theorems are formulated in terms of  purely
classical Hamiltonians $H^{(a)}(\beta,\mathbf{p},\mathbf{q})$  (\ref{Hi}) 
and classical trajectories, so that semiclassical partition
functions can be evaluated, e.g., via classical molecular dynamics
simulations. The Hamiltonians, however, contain additional potential-energy
terms, which are proportional to $\hbar^{2}$ and are temperature-dependent. 
The Hamiltonians $H^{(0)}(\beta,\mathbf{p},\mathbf{q})$ and  $H^{(1)}(\beta,\mathbf{p},\mathbf{q})$  
have already been used in classical, the so-called  Wigner-Kirkwood, molecular dynamics simulations (see Refs.   
\cite{ss,AlTi} and \cite{powles83,zoppi85,zoppi87}, correspondingly). Conceptually similar are also the 
Feynman-Hibbs molecular dynamics \cite{levesque04} and the quantized Hamilton dynamics \cite{prezhdo07}.
We mention also that the use of the path integral technique in dynamical computer simulations
 necessitates introducing additional fictitious  classical degrees of freedom into the system under study. 
For example, the ring polymer molecular dynamics is based on the classical evolution of the 
polymer beads  \cite{AlTi,ring,mano08}.  The path integrals have recently been applied to
the calculation of partition functions through the quantum work theorem  \cite{zon08a,zon08b}. 
In our semiclassical method,  no additional degrees of freedom are introduced. 
Furthermore, the higher order in $\hbar^{2}$ contributions can straightforwardly be 
incorporated into our main formulas (\ref{AB2}) and (\ref{Ja0}), if necessary. 
For example,  $H^{(1)}(\beta,\mathbf{p},\mathbf{q})$ 
evaluated up  to the terms of the order of  $\hbar^6$  has been used in Refs. \cite{zoppi85,zoppi87}.

The explicit form of the semiclassical work theorem
(Eqs. (\ref{AB2}) and (\ref{Ja0})) is not unique. In fact, we have 
derived four ($a=0,1,2,3$) different work theorems with different
definitions of the work operator $W^{(a)}$ (\ref{Ja1}). For each particular 
choice of the classical Hamiltonian $H^{(a)}(\beta,\mathbf{p},\mathbf{q})$,  
$W^{(a)}$ can definitely be treated as the  classical work, and 
 all $W^{(a)}$ coincide in the limit $\hbar \rightarrow 0$.
However, we cannot claim  any of $W^{(a)}$ to be a true semiclassical work operator. 
First, $W^{(a)}$ yield  different values of the work performed, 
and it is not clear which choice (if any) is preferable. 
Second, the mean value of the work operator $\overline  W^{(a)}$ (\ref{Wav})  does not generally coincide 
with the mean semiclassical energy difference $\Delta E$ (\ref{Eav}). 
This indicates, at the semiclassical level,  that it is hardly  possible to properly 
and unambiguously define the quantum work operator \cite{nie05,han07,lutz08,nol07,mahler07,kosJa,han09}. 
However, the ambiguity in the physical meaning of the quantity $W^{(a)}$ does 
not prevent us from using the semiclassical work theorem in 
practical calculations and classical molecular dynamics simulations, 
in order to get the system free energy differences or semiclassical averages.

Note finally that the present approach can be incorporated
into a general scheme developed in \cite{kosJa} to generate 
semiclassical versions of different fluctuation theorems for Hamiltonian systems in the equilibrium
and in a steady state. For example, treating the Hamiltonians (\ref{SB}) as true classical Hamiltonians, we 
can straightforwardly derive  the semiclassical counterpart of the Crooks transient fluctuation theorem 
\cite{cro99,cro00}: 
\begin{equation}
\left\langle\delta(w-W^{(a)}(\mathbf{p}_{0},\mathbf{q}_{0},t))\right\rangle_{0} =\left\langle\delta(w+W^{(a)}(\mathbf{p}_{t},\mathbf{q}_{t},t))\right\rangle _{t}\exp(\beta w)Z_{t}/Z_{0}.\label{Cr}\end{equation}
Here $W^{(a)}(\mathbf{p}_{0},\mathbf{q}_{0},t)$ and $W^{(a)}(\mathbf{p}_{t},\mathbf{q}_{t},t) $
 is the classical work (\ref{Ja1}) expressed as the 
function of the initial and final coordinates, correspondingly. 

\begin{acknowledgments}

This work has been supported by the Deutsche Forschungsgemeinschaft (DFG) through 
a research grant and through the DFG Cluster of Excellence ``Munich Centre of Advanced Photonics'' 
(MFG), and by American Chemical Society Petroleum 
Research Fund (44481-G6) (DSK). We are grateful to Michael Thoss and Wofgang Domcke for numerous useful discussions.   
 
\end{acknowledgments}

\newpage

\begin{figure}
\includegraphics[keepaspectratio,totalheight=12cm,angle=270]{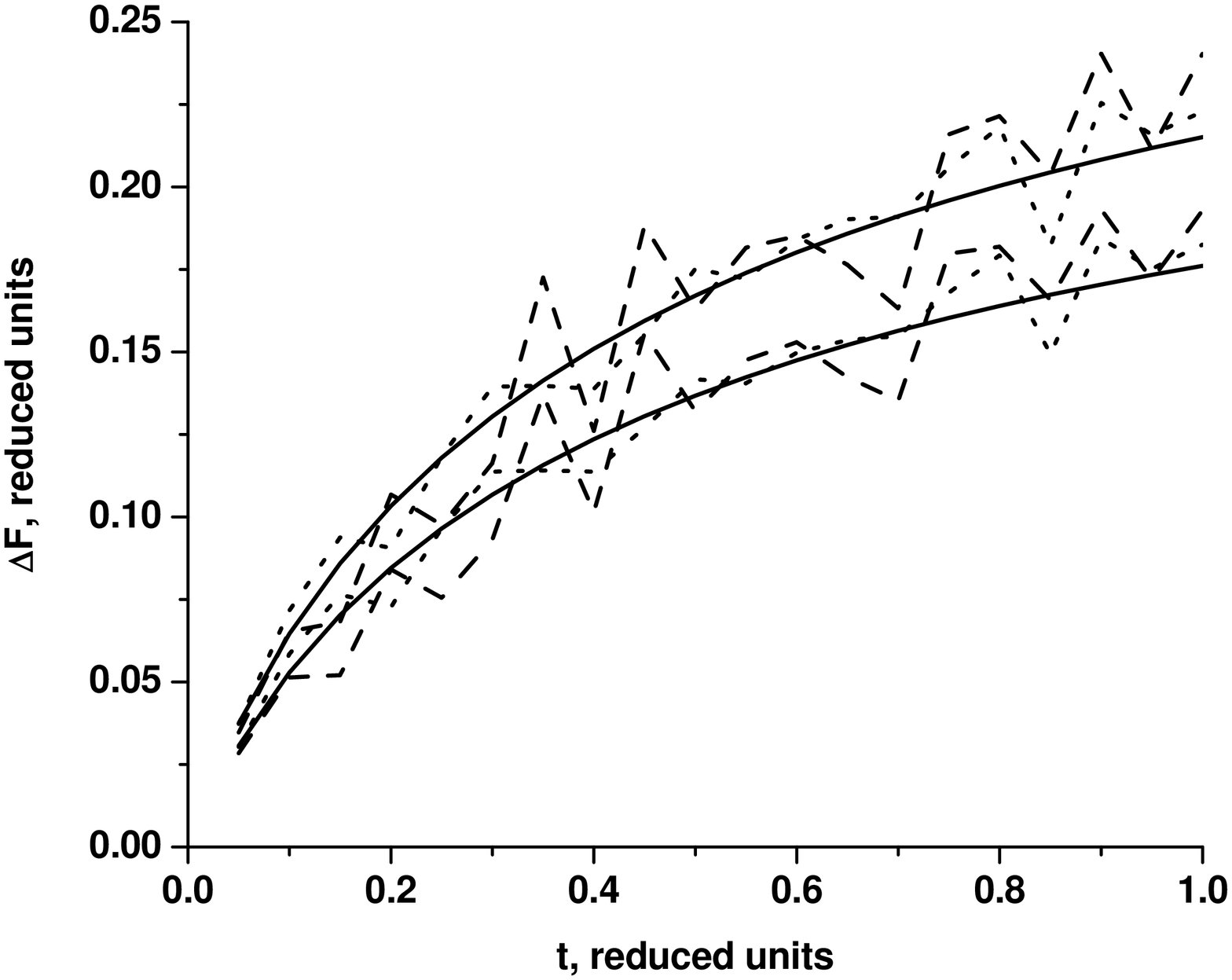}
\caption{
The difference between the free energies at times $t$ and $0$, 
$\Delta F = F_{t}-F_{0} = -\ln (Z_{t}/Z_{0})/ \beta$, as a function of $t$. The lower curves show the classical result ($h=0$) and the upper curves show the semiclassical result ($h=1$). Full lines correspond to "exact" semiclassical calculations, dashed and dotted lines correspond to "molecular dynamics" simulations (see text for details). 
}
\end{figure}


\begin{thebibliography}{10}
\bibitem{foot4}We can also interpret $H^{(0)}(\beta,\mathbf{p},\mathbf{q},t)$
as a true classical Hamiltonian. However, it contains mixed position-momenta 
terms and is thus infeasible for making calculations and/or simulations.
\bibitem{foot1}Definitely, the semiclassical theory does not work for $h=1$. We 
chose this value in our simulations just to get a bigger difference between the classical
 and semiclassical results.
\end{thebibliography}
\end{document}